\documentclass[runningheads]{llncs}
\usepackage[T1]{fontenc}
\usepackage{graphicx}
\usepackage{multirow}
\usepackage[utf8]{inputenc}
\usepackage{CJKutf8}
\usepackage[table]{xcolor}
\usepackage{tabularx}
\usepackage{float}
\usepackage{booktabs}
\usepackage{threeparttable}
\newcommand{\red}[1]{\textcolor{black}{#1}}

\begin{document}
\title{Towards Non-Latin Text and Layout Personalization for Enhanced Readability}
%
%

\author{Rina Buoy\inst{1,2}\orcidID{0000-0002-6960-4262} \and
Dylan berkamp Fouepe Dongmo\inst{2,4} \and
Vesal Khean\inst{3} \and
Simone Marinai\inst{4} \and
Koichi Kise\inst{2}}
\authorrunning{Buoy et al.}
%
\institute{Techo Startup Center, Ministry of Economy and Finance, Cambodia \email{rina.buoy@tsc.gov.kh}\and Osaka Metropolitan University, Japan \and Institute of Technology of Cambodia, Cambodia \and
University of Florence, Italy}
\maketitle              
\begin{abstract}
Reading has always been an integral part of both professional and personal life. Character and layout recognition and understanding by computers are well-explored areas. Nevertheless, how characters and layout are read and perceived by humans remains relatively underexplored. This work contributes to the field of human-document interaction (HDI) by investigating the effects of character and layout personalization on readability. The paper presents an empirical study on how parts-of-speech (POS)-based character and layout modifications can lead to overall improvements in both reading comprehension and memorization for two non-segmented, non-Latin writing systems: Khmer and Japanese. The experimental results from 43 participants suggest that, by bolding \red{POS-derived} content words, Khmer readers perform better on both reading comprehension and memorisation tasks, with a significant effect ($p$-values of 0.03 and 0.04, respectively). A similar overall tendency is also observed in a pilot study among Japanese readers (10 participants) using syntactic color-coding. In addition, the analyses of reading time, answering time, and perceived difficulty reveal that the proposed text styling technique does not increase any perceived difficulty, cognitive load, or reading effort for the Khmer readers. However, the Japanese readers experienced a decrease in reading speed. This study and its findings represent a significant step towards enabling dynamic, script-dependent personalization of character and layout to optimize human readability.

\end{abstract}
\section{Introduction}\label{intro}
Reading is an integral part of modern work and personal life; for many people, it is a primary means of acquiring information and knowledge. While computer-based reading is largely limited to character or word decoding, human reading is a complex, active, intent-driven cognitive process that converts visual signals into language, meaning, and comprehension~\cite{hoover_simple_1990}. In human reading, readability is a two-way notion, shaped by the reader’s skill and intent as well as by the material being read.   

Spaced or segmented text (i.e. adding between words or phrases) is believed to enhance human readability by optimizing word fixations~\cite{kobayashi_stepped_2015,hou_larger_2020}. However, many non-Latin scripts, such as Khmer and Japanese, do not use spaces between words, and adding artificial spaces to mark word boundaries may disrupt visual rhythm, slow reading, and cause discomfort~\cite{kobayashi_stepped_2015}. This raises a key question: how can we enhance the readability of non-segmented, non-Latin text without introducing spatial disruption? 

For Latin scripts, various experimental studies~\cite{hou_design_2022,dyson_influence_2001,li_adult_2019,bernard_page_2007} suggest that adjusting typographic and layout features can improve reading speed, comprehension, and user experience. Similarly, studies on non-segmented scripts (e.g. Chinese, Japanese, and Thai)~\cite{kobayashi_stepped_2015,oralova_effects_2021,hou_larger_2020,kohsom_adding_1997} show readability improvements by adjusting baseline layouts or forcing space boundaries. Despite growing interest, research on non-Latin scripts remains limited in both scale and scope, often relying on disruptive spatial interventions. 

To bridge this gap, we propose a modular part-of-speech (POS)-based text and layout styling technique that guides visual parsing without inserting spaces, targeting two non-segmented scripts: Khmer and Japanese. Unlike prior approaches, our technique applies high-resolution character-level styling using fine-grained POS cues to help readers distinguish syntactic roles (e.g., content vs. functional words). Crucially, we adapt the styling variable to the script's morphology: we apply an optimal \textit{bolding-based} style for Khmer, whereas we utilize a \textit{color-based} style for Japanese to prevent the visual crowding associated with bolding high-stroke-density Kanji.

To evaluate this approach, we conducted cross-lingual reading assessments. Participants read styled and non-styled texts in a randomly alternating order and complete reading comprehension and keyword memorization tasks. The experimental results from 43 Khmer participants demonstrate that bolding content words improves both comprehension and verbatim memorization with a significant effect ($p$-values of 0.03 and 0.04, respectively). Conversely, the analysis of a Japanese pilot cohort ($N=10$) revealed that while syntactic color-coding induced a slight reading time penalty (\red{$p=0.07$}), it significantly improved deep semantic comprehension and helped maintain accuracy during prolonged reading sessions. Nonetheless, it is important to note that the Japanese experiment as an exploratory pilot study ($N=10$) and the Khmer experiment ($N=43$) is the primary validation of the proposed method. Our core contributions can be summarized as follows:
\begin{enumerate}
    \item We propose a POS-based text styling technique for non-segmented, non-Latin scripts that styles words or phrases using fine-grained POS information.
    \item We devise readability experimental protocols and conducted two parallel experiments with Khmer and Japanese readers to evaluate the effectiveness of the proposed techniques.
    \item The experimental results show that applying the proposed technique improves Khmer readers’ performance on both reading comprehension and memorization tasks, with statistically significant effects. \red{A similar overall tendency was also observed among Japanese readers.}
\end{enumerate}


\section{Related Work}\label{relatedwork}

\subsubsection{Typographic Factors in Segmented Scripts}\label{relatedwork_latin}
Research on Latin scripts, where word boundaries are explicitly marked by spaces, has largely focused on optimizing typographic variables to enhance legibility. Arditi and Cho~\cite{arditi_serifs_2005} assessed the impact of typeface anatomy, finding that while serifs do not significantly alter reading speed for standard text, they improve legibility at smaller sizes or greater viewing distances. Beyond typeface selection, layout parameters such as font size and line spacing play a critical role in user experience. Hou et al.~\cite{hou_design_2022} highlighted that for elder adults reading on mobile devices, increased font size improves subjective satisfaction, although reading speed plateaus beyond a certain threshold.

Similarly, Dyson and Haselgrove~\cite{dyson_influence_2001} suggested that line length significantly impacts processing speed, recommending approximately 55 characters per line for optimal comprehension. Regarding vertical density, Bernard~\cite{bernard_page_2007} demonstrated that a substantial increase in line spacing (up to 178\%) could yield modest reading speed gains (26\%) for readers with simulated low vision. These studies collectively suggest that in segmented scripts, readability improvements are often sought through global layout adjustments rather than syntactic parsing aids.

\subsubsection{The Segmentation Challenge in \textit{Scriptura Continua}}\label{relatedwork_nonlatin}
In contrast to Latin scripts, languages utilizing \textit{scriptura continua}, such as Chinese, Japanese, Thai, and Khmer, rely on implicit context rather than explicit delimiters to define word boundaries. This absence of spacing imposes a distinct cognitive load, as readers must simultaneously decode characters and segment text into meaningful semantic units~\cite{kaing_towards_2021}.


Several studies have investigated the efficacy of inserting artificial spaces into these scripts. For Thai, Kohsom and Gobet~\cite{kohsom_adding_1997} provided evidence that adding inter-word spacing significantly improves both reading speed and comprehension. Similarly, for Chinese, Bai et al.~\cite{bai_reading_2008} and Oralova and Kuperman~\cite{oralova_effects_2021} found that while native readers implicitly group characters, the insertion of spaces at word boundaries can accelerate reading, whereas incorrect segmentation (e.g., at character boundaries) severely impairs performance.

However, the insertion of spaces is not universally beneficial. In the context of Japanese, Sainio et al.~\cite{sainio_role_2007} argued that while spaces might theoretically aid eye movements, they disrupt the familiar visual rhythm of the text, potentially causing discomfort. Consequently, researchers have explored alternative, non-spatial cues. Kobayashi et al.~\cite{kobayashi_stepped_2015} proposed a stepped-layout (indenting phrases vertically) to visualize boundaries without inserting whitespace. Their eye-tracking results indicated that such layout modifications could guide eye movements efficiently without imposing the additional cognitive load associated with artificial spacing.

\subsubsection{Gap Analysis}\label{gap_analysis}
Existing literature presents a dichotomy: segmented scripts benefit from typographic tuning, while non-segmented scripts struggle with boundary identification. Even if inserting spaces is a functional solution for some languages (Thai), it alters the native aesthetic and spatial density of others (Japanese). Crucially, neither standard typographic tuning nor artificial spacing conveys the \textit{syntactic structure} of the sentence (e.g., identifying the predicate or subject). 

To address this limitation, this paper proposes a POS-based styling technique. Instead of altering the spatial layout (which can be spatially disruptive), we leverage character-level attributes (boldness, color) to provide high-resolution syntactic information, thereby aiding human segmentation and parsing simultaneously. \red{The proposed method is considered less spatially disruptive, as it does not introduce unfamiliar or drastic layout changes to convey high-resolution syntactic information to readers.}

\section{Methodology}\label{methodology} 
\begin{figure}[t]
\centering
\includegraphics[width=0.9\textwidth]{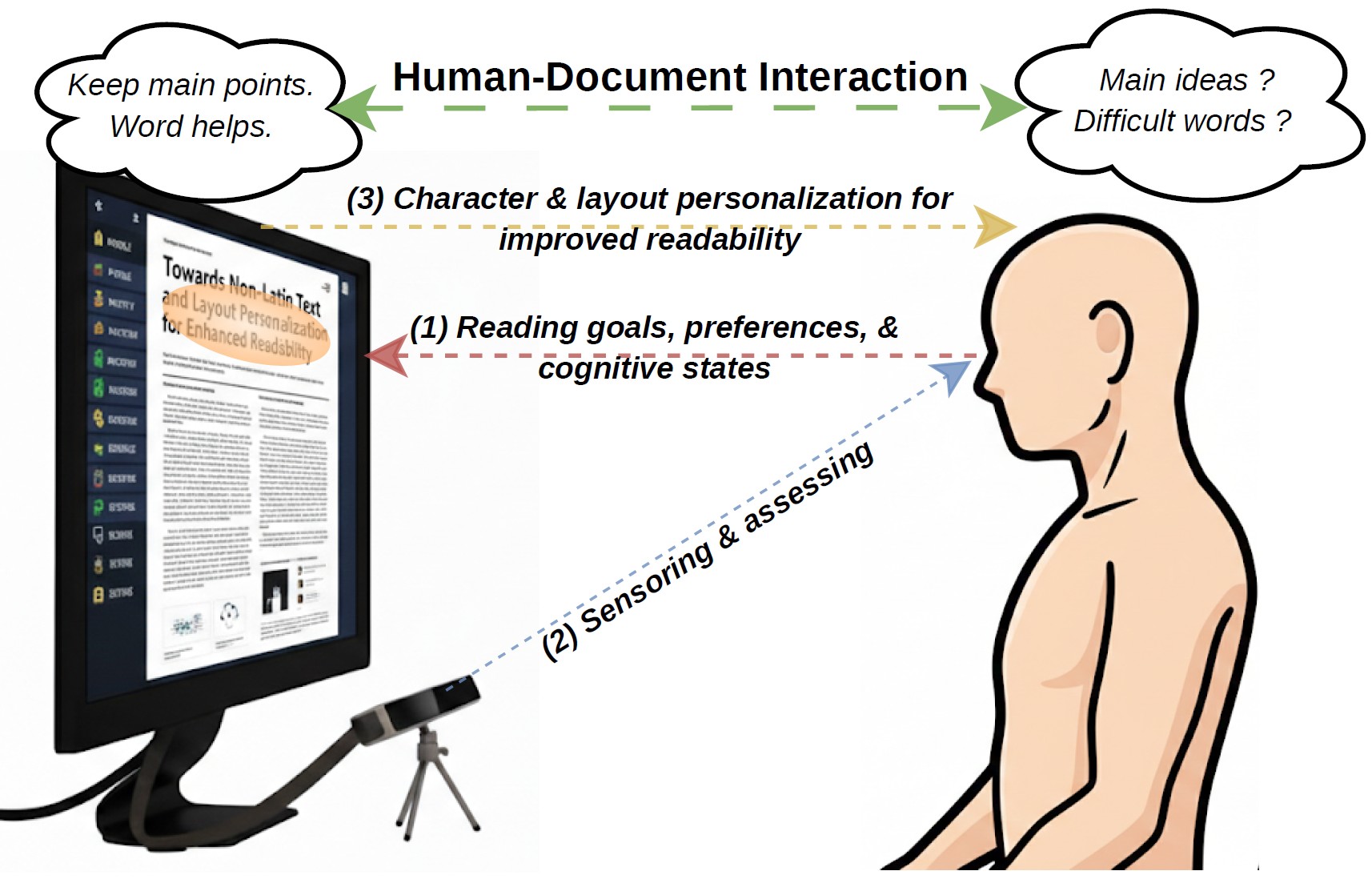}
\caption{\red{The ultimate framework for dynamic text and layout personalization for enhanced readability based on a reader's goals, preferences, and congitive states.}} \label{main_picture}
\end{figure}

\red{The ultimate goal, shown in Figure~\ref{main_picture}, is to build a dynamic system for personalizing character and layout that can enhance readers’ comprehension and performance, conditioned on a reader's goals, preferences, and congitive states. To achieve this goal, the system must continuously sense and assess a reader’ objectives, cognitive states, and preferences. Based on this information, the system can automatically adapt character and layout to maximize the reading reward, namely, readability. As an initial step toward achieving the ultimate goal described above, the proposed method is designed to measure the effects of altering the style of the text.}

\subsection{Proposed Method}\label{proposed_method} 

\begin{figure}[t]
\centering
\includegraphics[width=0.95\textwidth]{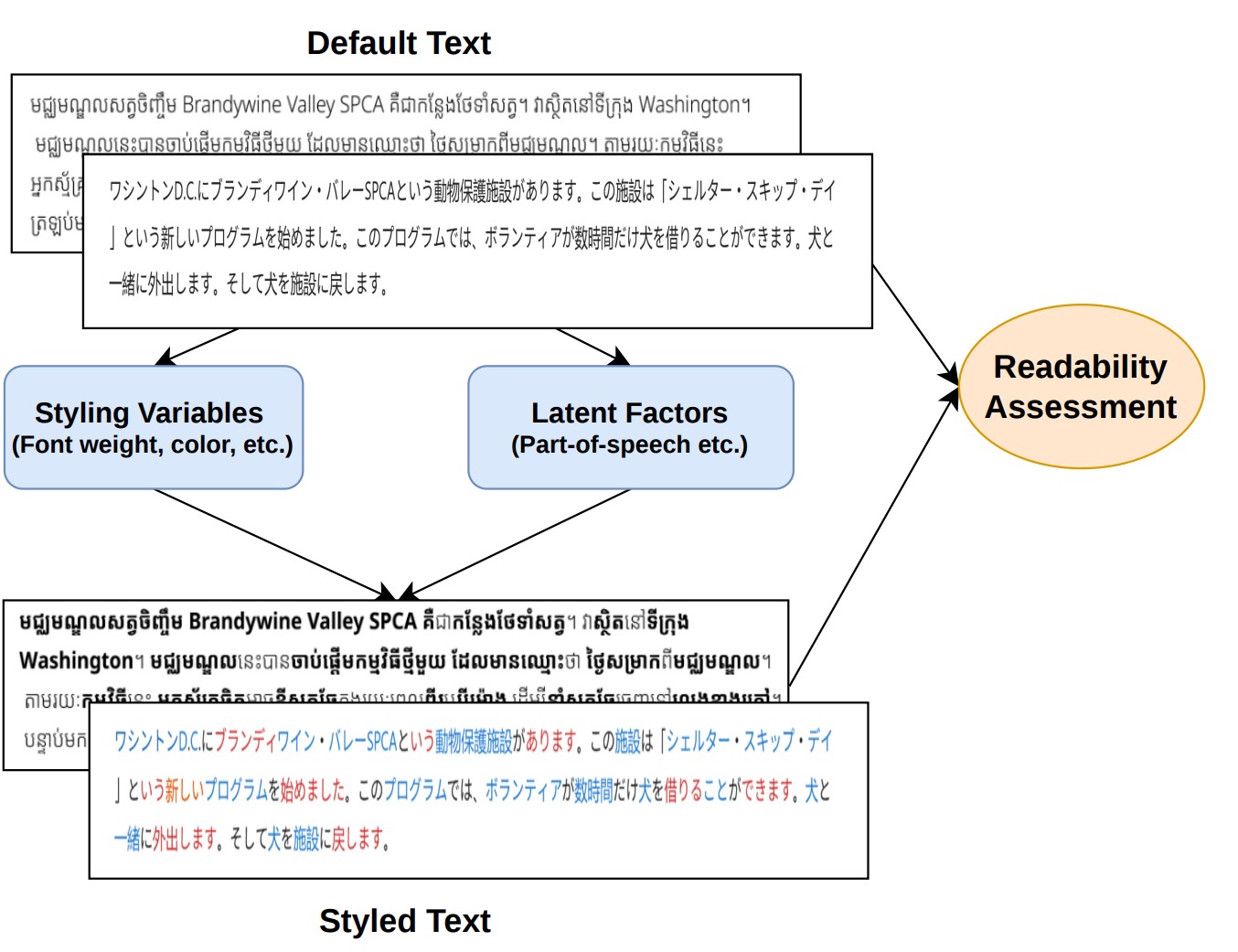}
\caption{The proposed method for POS-based text styling technique from a holistic view.} \label{overall_method}
\end{figure}

An overview of the proposed method is shown in Figure~\ref{overall_method} from a holistic perspective. As shown in the figure, the default (non-styled) text can be formatted according to a selected styling variable, such as font weight, typeface, colour, spacing, and others. The chosen styling variable is applied conditionally based on a selected latent factor, which may include \red{static linguistic features (e.g., POS, word boundaries)  and dynamic features (e.g., relevancy to a given theme)}. The impact of styling on reading performance is evaluated through a readability assessment test (Section~\ref{experiments}). \red{In the proposed method, one styling variable and one latent factor will be applied and evaluated at a time in order to independently and individually assess their respective impacts.}
 
\red{Based on the preliminary style preference surveys (Sections~\ref{khmer_preference_survey} and~\ref{japanese_heuristic}), we use font weight and colour as styling variables for Khmer and Japanese, respectively.}  In addition, POS is used as the latent factor, as it provides information about the syntactic roles of words in addition to indicating word boundaries. Furthermore, we developed a flexible text-formatting tool that conditionally applies various text styles based on POS information and other linguistic features; this tool is publicly available in an online repository\footnote{in review} and will be useful for future research in this field.

\subsection{POS-Based Khmer Text Styling}\label{khmer_pos_styling} 
Khmer is a highly analytic, non-segmented language with highly ambiguous POS and word boundaries~\cite{kaing_towards_2021}. Khmer readers rely heavily on contextual cues to implicitly form words and derive meaning as they read from left to right. Consequently, Kaing et al.~\cite{kaing_towards_2021} constructed a joint word segmentation and POS tagging dataset and proposed a unified modeling approach. The dataset includes major syntactic categories, such as noun ($n$), verb ($v$), adjective ($a$), modifier/complement ($o$), number ($1$), as well as other functional tags (e.g., determiners and articles).

\begin{figure}[t]
\centering
\includegraphics[width=0.9\textwidth]{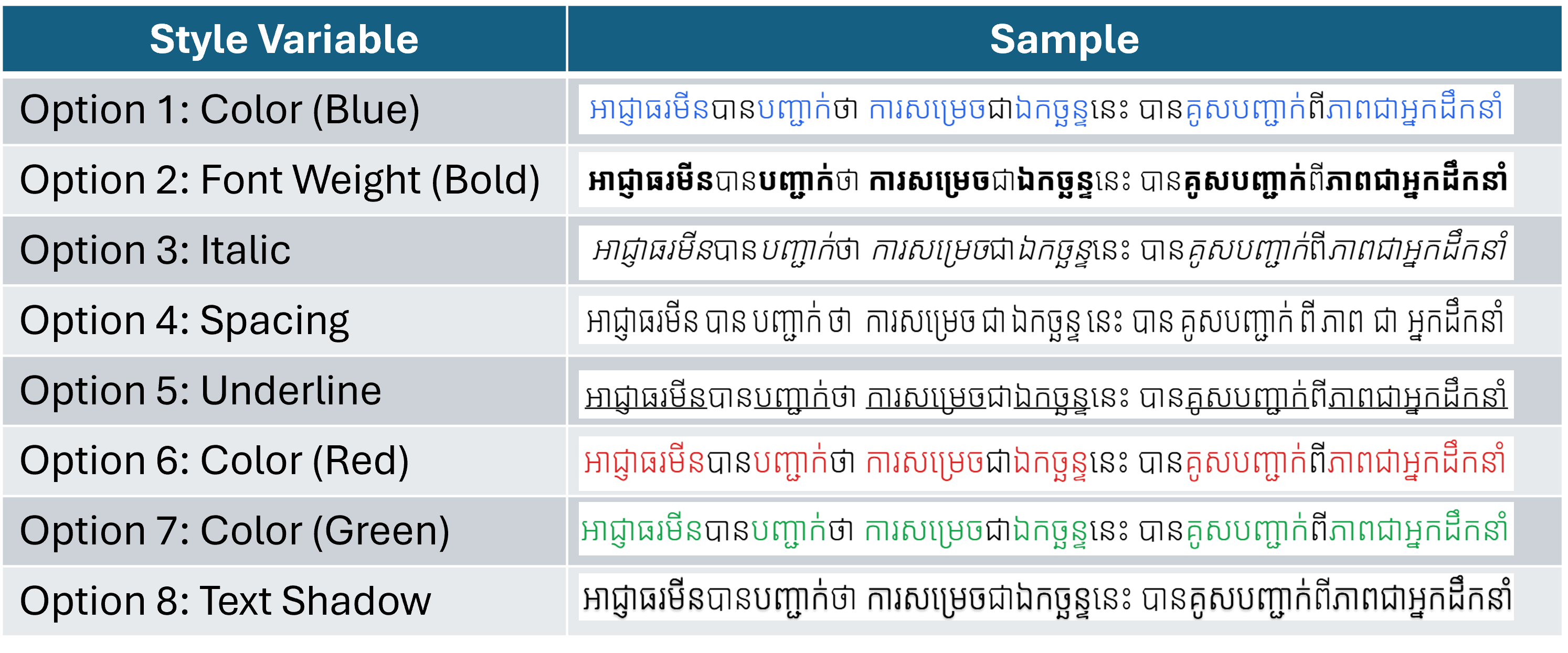}
\caption{Khmer style variables (\emph{English translation : the Mine Authority stated that this unanimous decision highlights Cambodia’s leadership}).} \label{style_preference}
\end{figure}

Consequently, we trained a Khmer Transformer~\cite{vaswani_attention_2017}-based model for joint word segmentation and POS tagging using the dataset from Kaing et al.~\cite{kaing_towards_2021}. Based on the predicted tags, words labeled as $n$, $v$, $a$, or $1$ are defined as content words, as they carry semantic meaning, while all other words are defined as functional words. A selected styling variable (e.g., font weight) is applied only to the content words to indicate both word boundaries and syntactic information. \red{Sample styled texts are provided in Figure~\ref{style_preference}.}

\subsection{\red{POS-Based Japanese Text Styling}}\label{japanese_pos_styling}

\begin{table}[t]
\caption{Syntactic Color Mapping Strategy for Japanese}
\label{tab:color_mapping}
\centering
\renewcommand{\arraystretch}{1.2}
\setlength{\tabcolsep}{4pt}
\resizebox{\textwidth}{!}{
\begin{tabular}{llcl}
\hline
\textbf{Grammatical Category} & \textbf{Syntactic Role} & \textbf{Color} & \textbf{Perceptual Justification} \\
\hline
NOUN, PROPN, PRON & Subjects, Entities & \cellcolor[HTML]{1976D2}\textcolor{white}{\textbf{Blue}} & Stable entity anchoring \\
VERB, AUX & Predicates, Actions & \cellcolor[HTML]{D32F2F}\textcolor{white}{\textbf{Red}} & Highlight semantic core \\
ADJ & States, Properties & \cellcolor[HTML]{E65100}\textcolor{white}{\textbf{Orange}} & Distinguish semantic modifiers \\
PART, ADP, SCONJ & Particles, Connectors & \cellcolor[HTML]{212121}\textcolor{white}{\textbf{Dark Gray}} & Reduce visual noise \\
PUNCT, SYM & Punctuation & \cellcolor[HTML]{AAAAAA}\textbf{Light Gray} & Minimize visual clutter \\
\hline
\end{tabular}%
}
\end{table}

We utilize GiNZA \cite{ginza} for syntactic analysis of Japanese, a \textit{scriptura continua} language mixing high-density logograms (Kanji) with phonetic syllabaries (Hiragana/Katakana). Built on the Universal Dependencies (UD) schema, GiNZA employs the SudachiPy tokenizer \cite{sudachipy}. We specifically opted for SudachiPy's \textbf{Mode C} (composite segmentation) to preserve compound nouns and idiomatic expressions as single, natural reading units, avoiding the over-segmentation common in traditional tools.

Unlike Khmer, where bolding provides optimal contrast, applying bold weights to high-stroke-density Kanji causes visual crowding and degrades legibility. Therefore, we employ syntactic color-coding. Cognitively, color enhances recognition memory \cite{wichmann2002contributions} and facilitates information chunking \cite{dzulkifli2013influence}. By mapping specific colors to grammatical roles (see Table \ref{tab:color_mapping}), our system leverages this chunking mechanism to visually segment the text. This approach reduces the extraneous cognitive load \red{associated with real-time word boundary resolution. It is important to notice that all selected colors were verified to meet the Web Content Accessibility Guidelines (WCAG) 2.1 Level AA standards, ensuring a contrast ratio of at least 4.5:1 against the white background.}

\section{Experiments}\label{experiments} 

\subsection{Khmer Style Preference Survey}\label{khmer_preference_survey} 

As shown in Figure~\ref{overall_method}, several styling variables can be applied to content words. However, the chosen styling variable must be carefully selected to ensure that word boundaries and syntactic information are conveyed to readers without introducing excessive unfamiliarity or discomfort and affecting reading attention and comprehension. To determine the most preferred style variable, we conducted an online preliminary style preference survey. Participants were asked to vote multiple pairs of style variables, shown in Figure~\ref{style_preference}, using an A/B comparison in terms of visual appeal, readability, and professionalism via a web-based survey. Each participant evaluated 28 distinct pairs of styled texts, one pair at a time, presented in random order. The style receiving the highest number of votes was selected for use in the final readability test.

\subsection{Japanese Style Validation (Heuristic)}\label{japanese_heuristic}
While the adoption of syntactic color-coding is primarily grounded in the cognitive principles and script morphology discussed in Section \ref{japanese_pos_styling}, we conducted a preliminary heuristic evaluation ($N=5$ native young adult speakers) to empirically validate this approach against other popular digital reading interventions, notably \textit{Bionic Reading}~\cite{bionic_reading}.

The evaluation confirmed our theoretical assumptions, showing a strong user preference for \textbf{Syntactic Color-Coding} (60\%) over Bionic Reading (27\%). This empirical validation confirmed Syntactic Color-Coding as the optimal intervention for the main readability assessment.

\subsection{Readability Assessment Protocol}\label{khmer_readability_survey} 

\begin{figure}
\centering
\fbox{\includegraphics[width=0.9\textwidth]{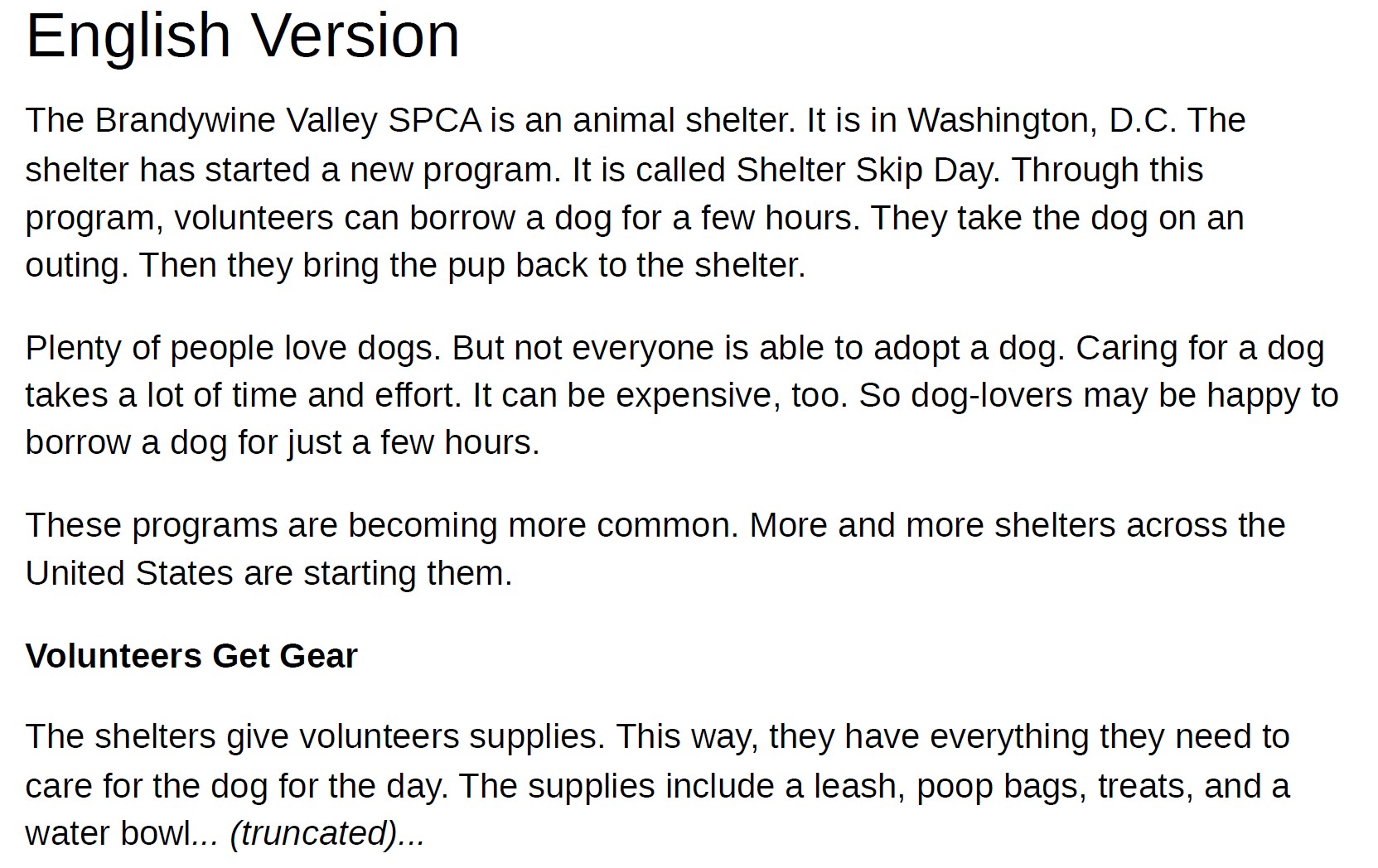}}
\caption{\red{A sample reading text in English.}} \label{sample_text}
\end{figure}

\begin{figure}
\centering
\includegraphics[width=0.9\textwidth]{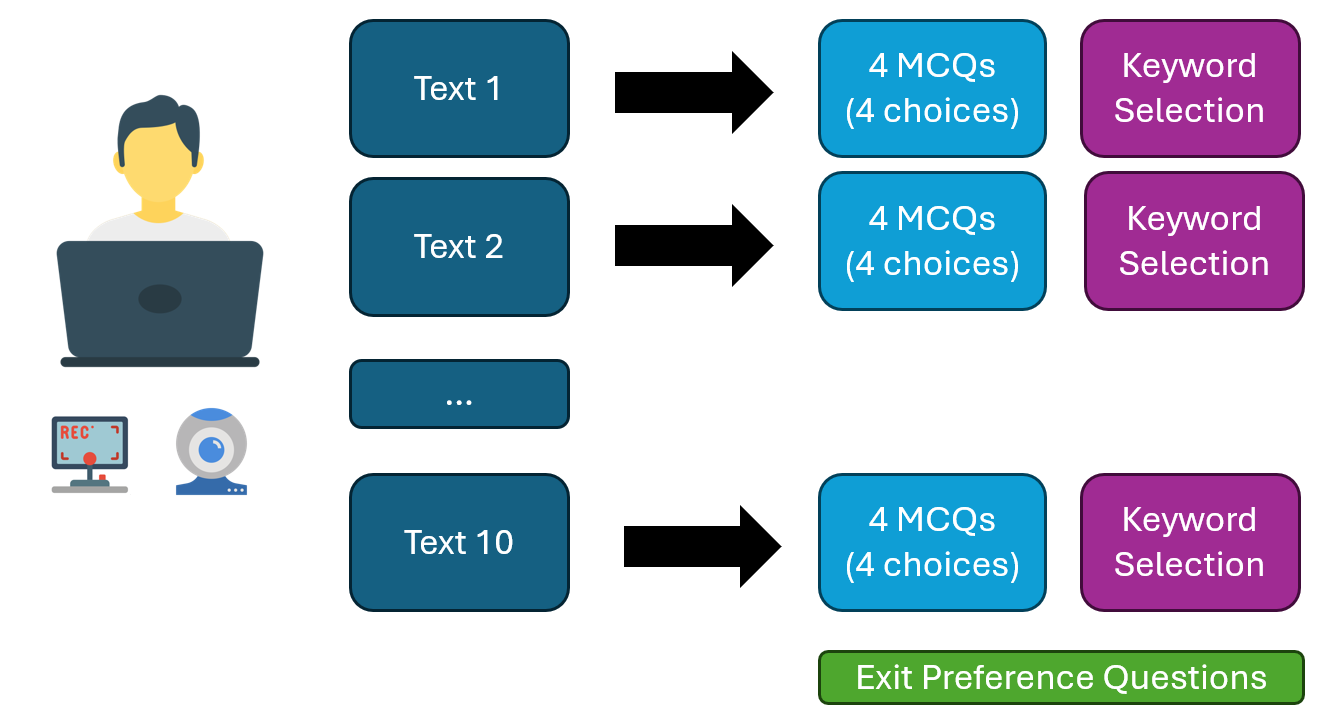}
\caption{The proposed readability assessment protocol.} \label{readability_test}
\end{figure}

\red{In this section, we discuss the readability assessment protocol that is applied to both Khmer and Japanese cohorts.} To enable cross-lingual comparisons of the proposed POS-based text styling technique, we collected 10 general-knowledge English articles of comparable length (approximately 250 words) and difficulty. \red{The general-domain and high-school-level articles were carefully selected from Newsela\footnote{https://newsela.com/} to avoid cultural or prior-knowledge bias across languages and to ensure an appropriate reading difficulty level for all participants. A sample text in English is provided in Figure~\ref{sample_text}}.  

\red{The articles were translated into Khmer and Japanese using the respective machine translation tools, followed by human curation to ensure translation quality.} Next, corresponding POS tags were assigned using the above respective POS tagging models. For each text, we generated two versions, styled and non-styled, containing identical content. Each article was accompanied by four multiple-choice questions (MCQs) for the reading comprehension task and \red{sample MCQ questions are provided in Table~\ref{tab:sample_question_types}}. Additionally, five exact keywords were extracted from each text, along with five similar but non-identical keywords, for the keyword selection task to assess memorization.


\red{The assessment protocol is} shown in Figure~\ref{readability_test}. Each participant is asked to read five styled and five non-styled texts in a randomized, alternating order. The sampled texts are non-duplicate and have comparable lengths and difficulty levels. Under this protocol, participants receive balanced exposure to styled and non-styled texts while minimizing potential \red{order effect}. For each text, participants answer four multiple-choice questions, including factual, inferential, global, and cloze questions, followed by a keyword selection task. Throughout the reading session, both screen recordings and webcam videos are captured.

\begin{table}[t]
\centering
\caption{\red{Sample comprehension MCQs.}}
\label{tab:sample_question_types}
\begin{tabular}{ll}
\toprule
\textbf{Question} & \textbf{Description} \\
\midrule
Q1 (Factual) &  \footnotesize What detail in the article shows that volunteers enjoy Shelter Skip Day?\\
Q2 (Inferential) & Which sentence explains why animal shelters want people to adopt dogs? \\
Q3 (Global) & What is the MAIN topic of the introduction? \\
Q4 (CLOZE) & Complete the sentence: ``Caring for a dog takes a lot of time and \underline{\hspace{1cm}}.'' \\
\bottomrule
\end{tabular}
\end{table}

\section{Results and Discussion}\label{result} 
\red{In the subsequent subsections, we begin by discussing the experimental results and key findings for Khmer, followed by Japanese.}
\subsection{Khmer Optimal Style}\label{khmer_optimal_style}

\begin{figure}[t]
\centering
\includegraphics[width=0.9\textwidth]{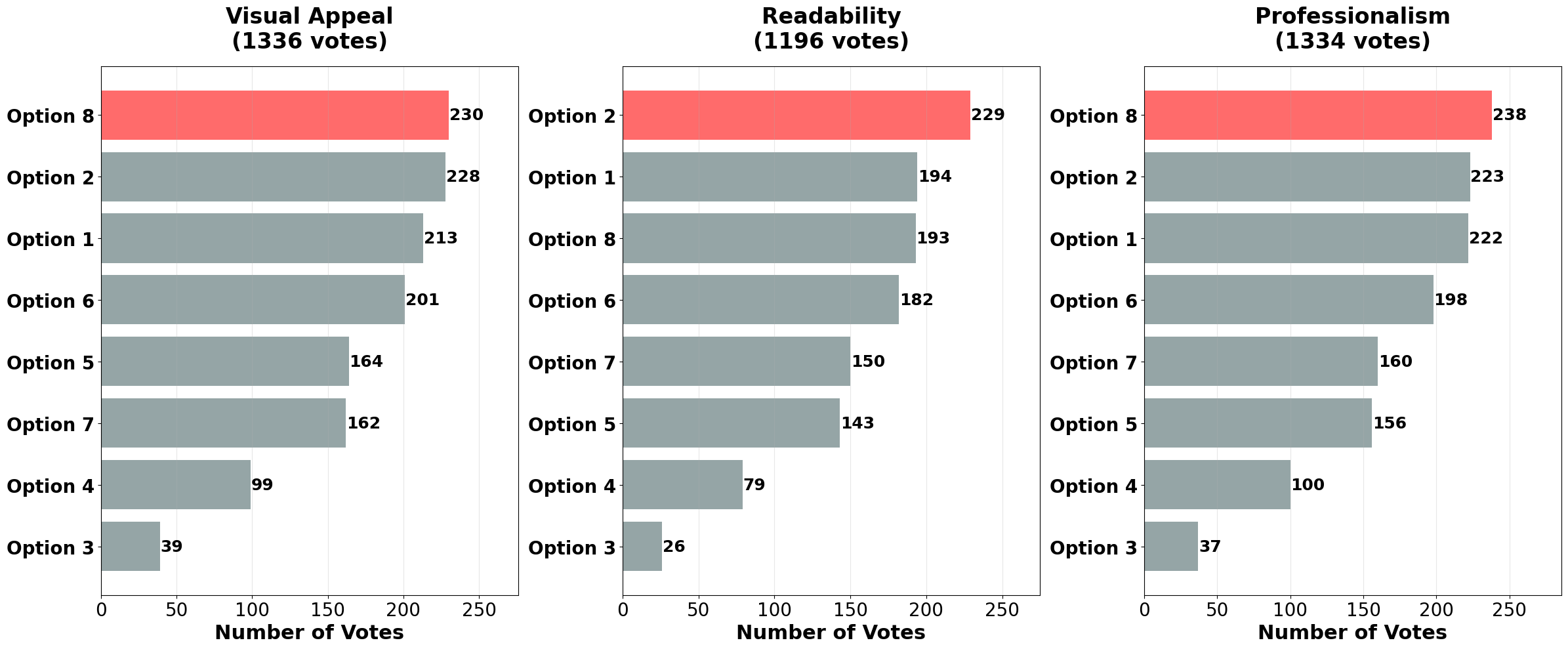}
\caption{Voting results of the Khmer style preference survey.} \label{style_preference_figure}
\end{figure}

\red{A total of 61 participants (university students) voluntarily took part in the Khmer style preference survey.} Participants who completed the survey too quickly (i.e., in less than five minutes) or took excessively long (i.e., more than 12 minutes) were excluded, as these behaviors indicated a lack of attention or potential connectivity issues, respectively. This resulted in 59 valid participants. The median completion time was approximately 8.02 minutes.

As shown in Figure~\ref{style_preference_figure}, \red{Option 2 (Font Weight – Bold)} emerged as the preferred choice, receiving the highest total votes across all dimensions \red{(i.e., visual appeal, readability, and professionalism)}, followed by \red{Option 8 (Text Shadow)}. Both of these leading styles enhance the visibility of content words through bolding, with Option 2 providing a higher contrast between content and functional words. Options 3 (Italic) and 4 (Spacing) were the least preferred by Khmer readers, as they were perceived as either visually uncomfortable or unfamiliar. Regarding the insertion of spacing, this finding aligns with observations \red{about inserting spaces} from previous studies~\cite{kobayashi_stepped_2015,sainio_role_2007} on Japanese script.

\subsection{Khmer Readability Assessment}\label{khmer_results}

\begin{figure}[t]
\centering
\includegraphics[width=0.55\textwidth]{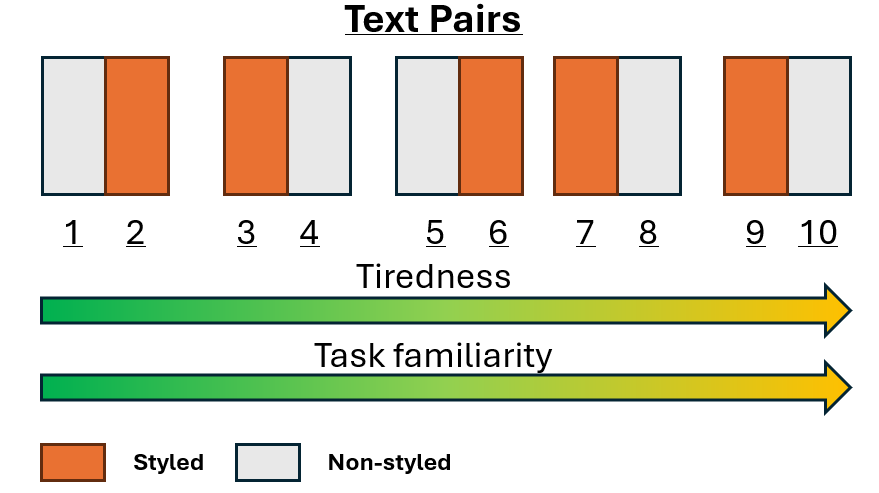}
\caption{An example text reading for a given participant.} \label{text_reading_order}
\end{figure}

\begin{table}[t]
\centering 
\caption{Average accuracies and differences  (\%) of styled vs. non-styled texts for the MCQ task. \textbf{S}: styled.  \textbf{NS}: non-styled. \textbf{Bold}: highest per question.}\label{mcq_results_accuracy}
\begin{tabular}{lccccccccc}
\hline
\multirow[b]{2}{*}{\textbf{ Text }} &
\multicolumn{3}{c}{\textbf{All Texts}} &
\multicolumn{3}{c}{\textbf{Texts 3--8}} &
\multicolumn{3}{c}{\textbf{Texts 5--8}} \\
\cline{2-4}\cline{5-7}\cline{8-10}
& \textbf{S} & \textbf{NS} & \textbf{$\Delta(S-NS)$}
& \textbf{S} & \textbf{NS}  & \textbf{$\Delta(S-NS)$}
& \textbf{S} & \textbf{NS}  & \textbf{$\Delta(S-NS)$}\\
\hline
\textbf{N} & 215 & 215 & 215 & 129 & 129 & 129 & 86 & 86 & 86\\

\textbf{Q1 (factual)} & 46.05 & 51.16 & \textbf{-5.12} & 45.74 & 53.49 & -7.75 & 44.19 & 54.65 & -10.47\\

\textbf{Q2 (inferential)} & 53.02 & 47.44 & 5.58 & 57.36 & 51.16 & \textbf{6.20} & 53.49 & 48.84 & 4.65 \\
\textbf{Q3 (global)} & 60.93 & 60.00 & 0.93 & 67.44 & 58.14 & \textbf{9.30} & 65.12 & 55.81 & \textbf{9.30}\\

\textbf{Q4 (cloze)} & 79.07 & 70.70 & 8.37 & 81.40 & 71.32 & 10.08 & 81.40 & 66.28 & \textbf{15.12}\\
\hline
\end{tabular}
\end{table}

\begin{table}[t]
\centering 
\caption{Question-level Chi-Squared statistical tests ($\chi^2$, $p$-value, and FDR-adjusted $p$-value in bracket) between styled vs. non-styled texts for the MCQ task. \textbf{Bold}: $p$-value<0.05.  \textbf{Italic}: $p$-value<0.1.}\label{mcq_results_test}
\begin{tabular}{lcccccc}
\hline
\multirow[b]{2}{*}{\textbf{ Text }} &
\multicolumn{2}{c}{\textbf{All Texts}} &
\multicolumn{2}{c}{\textbf{Texts 3--8}} &
\multicolumn{2}{c}{\textbf{Texts 5--8}} \\
\cline{2-3}\cline{4-5}\cline{6-7}
& \textbf{$\chi^2$} & \textbf{$p$-value}(\textbf{$q$-value})
& \textbf{$\chi^2$} & \textbf{$p$-value}(\textbf{$q$-value})  
& \textbf{$\chi^2$} & \textbf{$p$-value}(\textbf{$q$-value})\\
\hline

\textbf{Q1 (factual)}  & 0.93 & 0.33(0.33) & 1.26&0.26(0.33) & 1.49&0.22(0.33)\\

\textbf{Q2 (inferential)} & 1.13&0.29(0.57) & 0.77&0.38(0.57) & 0.21&0.65(0.65) \\
\textbf{Q3 (global)} & 0.01&0.92(0.92) & 2.01&0.16(0.41) & 1.19&0.27(0.41)\\

\textbf{Q4 (cloze)} & 3.57&\emph{0.06}(0.08) & 3.09&\emph{0.08}(0.08) & 4.33&\textbf{0.04}(0.08)\\
\hline
\end{tabular}
\end{table}

\begin{table}[t]
\centering 
\caption{Statistical paired-t test ($t$ test, $p$-value, and FDR-adjusted $p$-value in bracket) comparing styled vs. non-styled texts for the keyword selection task.  \textbf{S}: styled.  \textbf{NS}: non-styled.  \textbf{Bold}: $p$-value<0.05.}\label{keyword_selection_results}
\begin{tabular}{lcccccc }
\hline
\textbf{Text} & \textbf{N} & \multicolumn{2}{c}{\textbf{Accuracy, \%}} & \textbf{$\Delta(S-NS)$} & \textbf{$t$-test} & \textbf{$p$-value}(\textbf{$q$-value})  \\
\hline
\textbf{All texts } & 43 & 61.49& 58.22&3.28&   2.23 & \textbf{0.03}(0.03) \\
\hline
\textbf{Texts 3 to 8 } & 43 & 62.65 & 58.18 & 4.47 &   2.72 & \textbf{0.01}(0.02) \\
\hline
\textbf{Texts 5 to 8 } & 43& 62.64 & 56.77 & 5.87 &  2.68 & \textbf{0.01}(0.02) \\

\hline
\end{tabular}
\end{table}

\begin{table}[t]
\centering 
\caption{Reading, answering, and difficulty rating comparisons between the styled and non-styled texts. \textbf{S}: styled.  \textbf{NS}: non-styled.  \textbf{Bold}: $p$-value<0.05.  \textbf{Italic}: $p$-value<0.1.}\label{others_results_test}
\begin{tabular}{lcccccc}
\hline
\textbf{Measure} & \textbf{N} & \textbf{Mean (S)}& \textbf{Mean (NS)} & \textbf{$\Delta(S-NS)$} & \textbf{$t$-test} & \textbf{$p$-value}  \\
\hline
\textbf{Reading Time (sec.) } & 43 & 107.59 &  109.86 & -2.27 & -1.49 & 0.14\\
\textbf{Answering Time (sec.) } & 43 & 93.63 &  93.19 &  0.44  & -0.15  & 0.88\\
\textbf{Difficulty Rating} & 43 & 2.95 & 3.00 & -0.04 &  -0.66&    0.52 \\

\hline
\end{tabular}
\end{table}

50 university students participated in the readability assessment test using a web-based survey application on a laptop computer equipped with a webcam in a classroom setting. Of these, 44 participants completed the experiment by reading and answering all 10 texts, while six participants partially completed the experiment due to technical issues, such as poor internet connectivity, browser-related problems, or disk failures.

The median completion time was approximately 40.66 minutes. One participant with a completion time of less than 30 minutes was excluded, resulting in a total of 43 valid participants. In terms of demographics, 53.5\% of participants were female, and 83.7\% were undergraduate students aged between 18 and 24. All participants were native speakers of Khmer.

Before analyzing the experimental results, it is important to reiterate that each participant read five styled and five non-styled texts (i.e., five pairs) in a randomized, alternating order, an example of which is shown in Figure~\ref{text_reading_order}. As illustrated in the figure, two factors must be considered when interpreting the results: fatigue and task familiarity. Participants may initially require time to become comfortable with styled texts. However, as their familiarity with the styling increases, fatigue from the reading session may also accumulate. These competing factors should be taken into account during analysis. In this regard, we grouped the texts into three categories for subsequent analyses: \textbf{All Texts}, which includes all texts; \red{\textbf{Texts 3–8}, with texts from 3 to 8 and  \textbf{Texts 5–8}, with texts from 5 to 8. These subsets attempt to focus on experiments where the reader is familiar with the task and not too tired.}

We begin by analyzing whether there is a significant difference in accuracy between the styled and non-styled texts for the MCQ task. In other words, we test whether the distribution of correct and incorrect answers differs significantly between styled and non-styled texts for each question type across all participants. Table~\ref{mcq_results_accuracy} presents the average accuracy comparisons between styled and non-styled texts, grouped by question type and text groups. The table shows that, except for factual questions, styled texts resulted in higher accuracy than non-styled texts across all question types. These accuracy improvements remain consistent even when excluding early and late text pairs (i.e., \textbf{Texts 3–8} and \textbf{Texts 5–8}), as participants may be less familiar with styling at the beginning or fatigued toward the end of the session. Specifically, styling led to maximum average accuracy improvements of 15.12\% for cloze questions, 9.3\% for global questions, and 6.2\% for inferential questions. In contrast, styling decreased performance for factual questions.

Furthermore, as shown in Table~\ref{mcq_results_test}, the accuracy improvement for cloze questions was statistically significant ($p$-value < 0.05) when considering only \textbf{Texts 5–8}, and marginally significant ($p$-value < 0.1 and $q$-value < 0.1) in other cases and with Benjamini-Hochberg (BH) false discovery rate (FDR) correction, respectively. These results suggest that the proposed POS-based text styling technique can enhance readers’ overall comprehension and ability to recall specific and key information. However, the method is less effective when the reading goal is to retrieve specific factual details. This is because the proposed method operates at the syntactic level and does not explicitly encode semantic salience. Consequently, words that carry essential factual information are not preferentially highlighted.

For the keyword selection task, the goal was to determine whether the difference in average accuracy between styled and non-styled texts was significant. Paired t-tests were conducted for different text groups, as shown in Table~\ref{keyword_selection_results}. The table indicates that the average accuracies of styled texts were higher than those of non-styled texts across all text groups, and these improvements were statistically significant ($p$-value < 0.05) even with FDR correction ($q$-value < 0.05). These results suggest that the proposed POS-based text styling technique can aid readers in recalling specific parts of the text.

Regarding temporal and cognitive metrics, Table~\ref{others_results_test} demonstrates that the proposed text styling introduces no significant overhead. Reading speed, answering time, and perceived difficulty remained stable ($p$-values > 0.1). This confirms that POS-based bolding in Khmer successfully guides semantic parsing without imposing an extraneous cognitive load or spatial disruption.

In summary, the proposed POS-based text styling technique can enhance reading performance for both comprehension and keyword selection tasks without increasing perceived difficulty, cognitive load, or reading effort.

\subsection{\red{Japanese Readability Assessment}}\label{japanese_results}

\begin{table}[t]
\centering 
\caption{Average accuracies and differences (\%) of styled vs. non-styled texts for the Japanese MCQ task. \textbf{S}: styled. \textbf{NS}: non-styled. \textbf{Bold}: maximum variations.}
\label{tab:jap_mcq_accuracy}
\resizebox{\textwidth}{!}{%
\begin{tabular}{lccccccccc}
\hline
\multirow[b]{2}{*}{\textbf{Question Type}} &
\multicolumn{3}{c}{\textbf{All Texts}} &
\multicolumn{3}{c}{\textbf{\red{Texts 3--8}}} &
\multicolumn{3}{c}{\textbf{\red{Texts 5--8}}} \\
\cline{2-4}\cline{5-7}\cline{8-10}
& \textbf{S} & \textbf{NS} & \textbf{$\Delta(S-NS)$}
& \textbf{S} & \textbf{NS}  & \textbf{$\Delta(S-NS)$}
& \textbf{S} & \textbf{NS}  & \textbf{$\Delta(S-NS)$}\\
\hline
\textbf{Q1 (Factual)}     & 73.68 & 64.91 & 8.77 & 87.88 & 71.43 & 16.45 & 80.95 & 78.26 & 2.69 \\
\textbf{Q2 (Inferential)} & 75.44 & 70.18 & 5.26 & 84.85 & 65.71 & \textbf{19.13} & 85.71 & 69.57 & 16.15 \\
\textbf{Q3 (Global)}      & 87.72 & 80.70 & 7.02 & 90.91 & 74.29 & 16.62 & \textbf{100.00} & 73.91 & \textbf{26.09} \\
\textbf{Q4 (Cloze)}       & 71.93 & 77.19 & -5.26 & 66.67 & 80.00 & \textbf{-13.33} & 80.95 & 82.61 & -1.66 \\
\hline
\end{tabular}%
}
\end{table}

\begin{table*}[t]

\centering 

\begin{threeparttable}
\caption{GEE modeling the odds of correct answers (Styled vs. Non Styled texts). Odds Ratios (OR) > 1 indicate styling improves accuracy. (\textbf{$q$-value}): FDR-adjusted $p$-value. \textbf{Bold}: $p$-value < 0.05. \emph{Italic}: $p$-value < 0.1.}
\label{tab:gee_results}
\begin{tabular}{lcccccc}
\hline
\multirow[b]{2}{*}{\textbf{Question Type}} &
\multicolumn{2}{c}{\textbf{All Texts}} &
\multicolumn{2}{c}{\textbf{\red{Texts 3--8}}} &
\multicolumn{2}{c}{\textbf{\red{Texts 5--8}}} \\
\cline{2-3}\cline{4-5}\cline{6-7}
& \textbf{OR} & \textbf{$p$-value}(\textbf{$q$-value}) 
& \textbf{OR} & \textbf{$p$-value}(\textbf{$q$-value})  
& \textbf{OR} & \textbf{$p$-value}(\textbf{$q$-value}) \\
\hline
\textbf{Q1 (Factual)}     & 1.51 & 0.43(0.65) & 2.90 & \emph{0.10}(0.29) & 1.18 & 0.81(0.81) \\
\textbf{Q2 (Inferential)} & 1.31 & \textbf{0.02}(0.05) & 2.92 & \textbf{0.03}(0.05) & 2.63 & 0.16(0.16) \\
\textbf{Q3 (Global)}      & 1.71 & 0.20(0.20) & 3.46 & \textbf{0.03}(0.04) & $\infty$* & \textbf{<0.01}(0.01) \\
\textbf{Q4 (Cloze)}       & 0.76 & 0.43(0.64) & 0.50 & 0.20(0.60) & 0.89 & 0.87(0.86) \\
\hline
\end{tabular}
\begin{tablenotes}
        \scriptsize
        \item[*] Infinite OR due to complete statistical separation (100\% accuracy in the styled condition).
    \end{tablenotes}
\end{threeparttable}
\end{table*}

\begin{table}[t]
\centering 
\caption{Generalized Linear Mixed Models (GLMM) for Reading Time, Keyword Accuracy, and Group Consensus (Crowd Truth) across temporal phases. \red{\textbf{S}: styled vs. \textbf{NS}: non-styled Texts.} \textbf{Bold}: $p$-value < 0.05. \emph{Italic}: $p$-value < 0.1.}
\label{tab:jap_glmm_combined}
\resizebox{\textwidth}{!}{%
\begin{tabular}{llcccc}
\hline
\textbf{Metric} & \textbf{Phase} & \textbf{Mean (\red{\textbf{S}})} & \textbf{Mean \red{\textbf{(NS)}}} & \textbf{Coeff} \red{ $\Delta$} & \textbf{$p$-value} \\
\hline
\multirow{3}{*}{\textbf{Reading Time (sec.)}} 
& \red{All Texts} & 94.20 & 86.45 & +8.21 & \emph{0.07} \\
& Texts 3--8 & 95.46 & 85.96 & +9.50 & \emph{0.10} \\
& Texts 5--8 & 94.27 & 87.00 & +7.27 & 0.38 \\
\hline
\multirow{2}{*}{\textbf{Keyword Accuracy (\%)}} 
& \red{All Texts} & 72.98 & 74.47 & -1.05 & 0.79 \\
& Texts 5--8 & 73.81 & 81.52 & -6.65 & 0.20 \\
\hline
\multirow{2}{*}{\textbf{Group Consensus (\%)}} 
& \red{All Texts} & 77.89 & 80.00 & -1.19 & 0.66 \\
& Texts 5--8 & 83.48 & 76.19 & -6.41 & 0.16 \\
\hline
\end{tabular}%
}
\end{table}

\begin{figure}[t]
    \centering
    \includegraphics[width=0.9\textwidth]{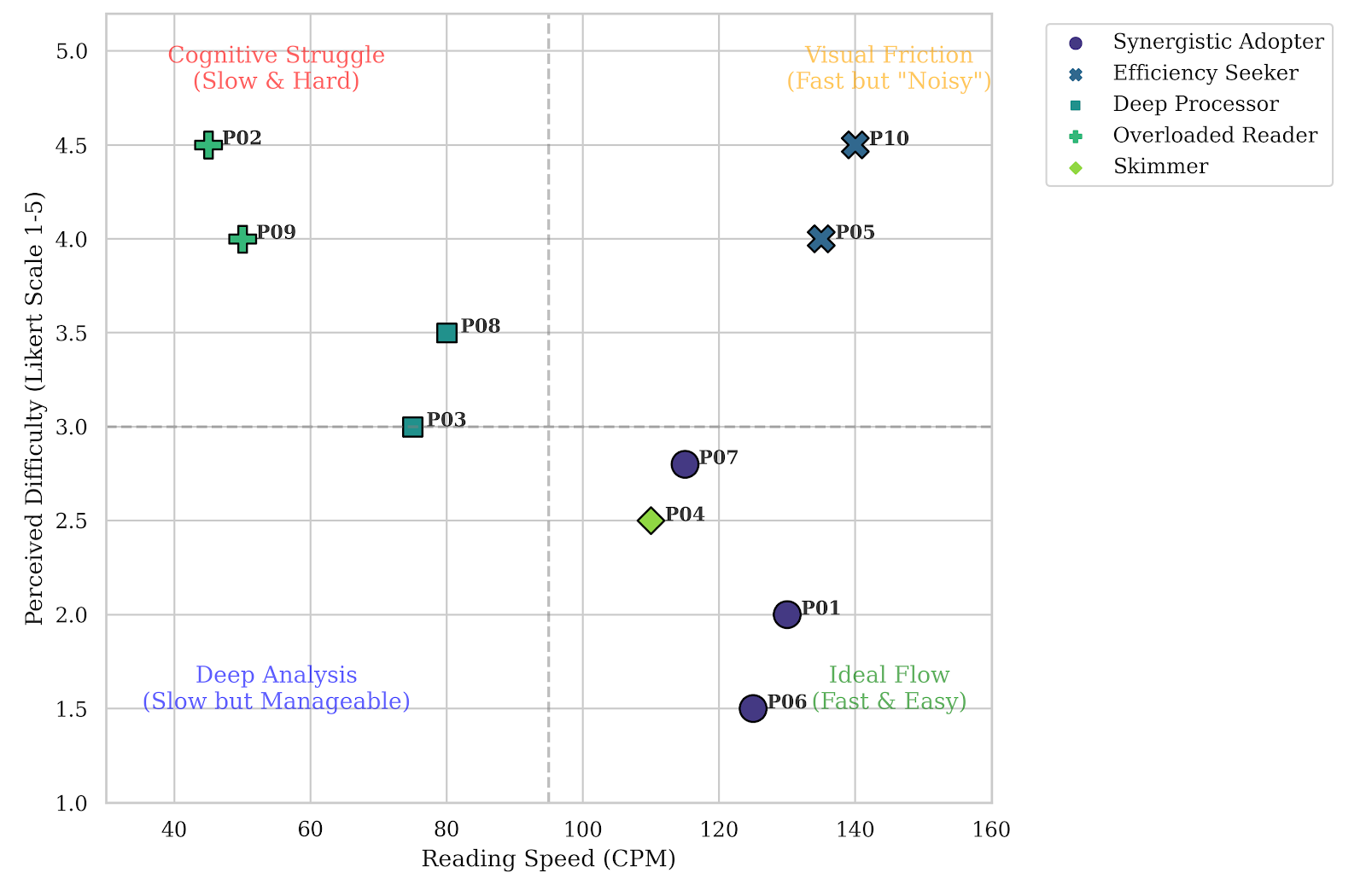}
    
    \caption{Conceptual scatter plot of reader profiles based on Reading Speed vs. Perceived Difficulty.}
    \label{fig:profiles}
\end{figure}

\subsubsection{Speed-Accuracy Trade-off and Temporal Dynamics}
A pilot cohort of 10 native Japanese speakers participated in the assessment. Applying the same temporal grouping protocol established in Section~\ref{khmer_results}, \red{we utilized Generalized Linear Mixed Models (GLMM) to account for random effects arising from individual subject variance in this repeated-measures design.}

As shown in Table~\ref{tab:jap_glmm_combined}, the time penalty incurred by the syntactic styling was minimal. Across all texts, the styling added a marginally significant average of 8.21 seconds per text ($p=0.07$). 
Despite this minor deceleration, raw accuracy metrics (Table~\ref{tab:jap_mcq_accuracy}) and the subsequent GEE model (Table~\ref{tab:gee_results}) reveal a trade-off. During the \textit{Familiarity Phase} (Texts 3--8), styling provided significant benefits, increasing the odds of answering inferential questions correctly (OR = 2.92) and improving global comprehension odds by 3.46. However, the odds for Cloze questions decreased (OR = 0.50), suggesting that while color-coding aids semantic understanding, it may interfere with the grammatical fluidity required for fill-in-the-blanks tasks.

Crucially, during the \textit{Fatigue Phase} (Texts 5--8), the styled condition showed a strong protective effect. While baseline accuracy declined in the non-styled group, the GEE model exhibited complete statistical separation ($p<0.01$) for styled global questions, maintaining high accuracy rates.

\subsubsection{Semantic Stability and Reader Profiles}
To evaluate whether layout personalization altered keyword memorization, \red{we modeled Keyword Accuracy and Crowd Truth consensus \cite{aroyo2014three} (Table~\ref{tab:jap_glmm_combined}). Unlike binary majority voting which treats disagreement as error, Crowd Truth harnesses inter-rater disagreement to capture continuous semantic ambiguity.} The analysis confirmed that color-coding maintained stability in both isolated keyword recall ($p>0.20$) and group consensus ($p>0.16$). The readers converged on the same semantic heavyweights, indicating the design is semantically non-destructive.

By cross-referencing the reading speed in Characters Per Minute (CPM) with perceived difficulty, we identified distinct reader profiles summarized in Figure~\ref{fig:profiles}: \textit{Synergistic Adopters} (stable speed, reduced difficulty), \textit{Efficiency Seekers} (rushed reading), \textit{Deep Processors} (slowdown for higher accuracy), \textit{Overloaded Readers} (slow speed, high difficulty), and \textit{Skimmers} (rapid scanning with moderate difficulty).

\subsubsection{Cross-Lingual Synthesis}
Comparing both tracks highlights how script morphology dictates HDI strategies. In the Khmer cohort, applying \textit{Syntactic Bolding} to content words significantly improved isolated keyword selection ($p < 0.05$). Conversely, for the high stroke density of Japanese Kanji, \textit{Syntactic Color-Coding} functioned less as a memorization tool and more as an attention stabilizer, optimizing deep semantic parsing and resilience against fatigue.

Nonetheless, we treat the Japanese experiment as an exploratory pilot study and position the Khmer experiment as the primary validation of the proposed method.

\section{Limitations and Future Work}
We identify the following limitations associated with this study and propose directions for future research:

\begin{enumerate}
    \item \textbf{Data Granularity:} The current reading assessment, relying on MCQ comprehension and keyword selection tasks, provides only low-resolution data on reading behavior.  Future experiments should incorporate high-resolution sensored data, such as eye-tracking, to assess reading performance, effort, and cognitive states. 

    \item \textbf{Adaptive Styling:} \red{In this study, a single style variable (e.g., bolding or coloring) was applied uniformly to all participants. Future work will focus on developing an adaptive styling technique that can adjust the styling intensity according to a reader’s preferred comfort level.}
    

    \item \textbf{Dynamic Styling:} \red{In this study, styling is applied to POS-derived content words, and POS is treated as a static linguistic latent feature. Future work will focus on deriving content words dynamically with respect to a reader’s preferences and intents.}
    
    
\end{enumerate}

\section{Conclusions}
In this study, we proposed a modular POS-based text and layout styling for non-segmented scripts, specifically Khmer and Japanese. Unlike traditional methods that insert artificial whitespace, our approach enhances readability by clarifying syntactic boundaries without introducing spatial disruption.

Experimental results from 43 Khmer readers demonstrated that bolding POS-derived content words significantly improved reading performance, yielding maximum accuracy gains of 15.12\% for a comprehension task and significant improvements in a keyword selection task ($p < 0.05$) without increasing perceived difficulty, cognitive load, or reading effort. Conversely, the pilot study with 10 Japanese readers revealed that high-stroke-density scripts respond differently. Using syntactic color-coding, readers adopted a deep processing behavior. While incurring a minimal time penalty ($p=0.074$), the styling yielded significant cognitive benefits during the familiarity phase, increasing the odds of deep semantic comprehension. Crucially, as fatigue accumulated, the color-coding mitigated performance loss, maintaining high accuracy rates while preserving natural group consensus on semantic salience.

These divergent findings suggest that readability enhancement is dictated by script morphology and cognitive constraints. While selective bolding optimizes information retrieval in moderately dense scripts (Khmer), holistic color-coding facilitates structural processing and fatigue resilience in logographic scripts (Japanese). \red{This research is a step towards realzing a dynamic text and layout personalization system for enhanced readability based on a reader's goals, preferences, and congitive states.}

%
%
%
%
\bibliographystyle{splncs04}

\end{document}